\documentclass[sigconf,natbib=false,nonacm]{acmart}

\renewcommand\footnotetextcopyrightpermission[1]{} 
\usepackage{enumitem}

\usepackage{amssymb}
\usepackage{algorithm}
\usepackage{algpseudocode}
\algrenewcommand{\algorithmiccomment}[1]{\hfill\textcolor{black!60}{$\triangleright$ \small \textit{#1}}}
\mathchardef\mhyphen="2D 

\usepackage{booktabs}

\usepackage{float}
\usepackage{listings}

\newfloat{lstfloat}{tbp}{lop}
\floatname{lstfloat}{Listing}

\usepackage{multirow}
\usepackage{multicol}
\usepackage{makecell}

\newlist{todolist}{itemize}{2}
\setlist[todolist]{label=$\square$}
\usepackage{pifont}

\usepackage{hyperref}
\hypersetup{
 unicode=true,
 pdfstartview={FitH},
 colorlinks=true,
 citecolor=blue,
 linkcolor=blue,
 urlcolor=blue
}

\usepackage[normalem]{ulem}

\RequirePackage[
  minbibnames=6,
  maxbibnames=6,
  ]{biblatex}
\AtBeginDocument{%
  }

\copyrightyear{2023}
\acmYear{2023}
\setcopyright{none}
\acmBooktitle{28th Asia and South Pacific Design Automation Conference (ASPDAC '23), January 16--19, 2023, Tokyo, Japan}
\acmPrice{15.00}
\acmDOI{10.1145/3566097.3567892}
\acmISBN{978-1-4503-9783-4/23/01}

\begin{document}
\newcommand{\schedscale}{0.25}
\newcommand{\tabscale}{0.85}

\title[Iris: Automatic Generation of Efficient Data Layouts for High Bandwidth Utilization]{Iris: Automatic Generation of Efficient Data Layouts\\ for High Bandwidth Utilization}

\author{Stephanie Soldavini}
\affiliation{%
 \institution{Politecnico di Milano}
 \city{Milan}
 \country{Italy}
}
\email{stephanie.soldavini@polimi.it}
\author{Donatella Sciuto}
\affiliation{%
 \institution{Politecnico di Milano}
 \city{Milan}
 \country{Italy}
}
\email{donatella.sciuto@polimi.it}
\author{Christian Pilato}
\affiliation{%
 \institution{Politecnico di Milano}
 \city{Milan}
 \country{Italy}
}
\email{christian.pilato@polimi.it}



\begin{abstract}
Optimizing data movements is becoming one of the biggest challenges in heterogeneous computing to cope with data deluge and, consequently, big data applications. When creating specialized accelerators, modern high-level synthesis (HLS) tools are increasingly efficient in optimizing the computational aspects, but data transfers have not been adequately improved. 
To combat this, novel architectures such as High-Bandwidth Memory with wider data busses have been developed so that more data can be transferred in parallel. 
Designers must tailor their hardware/software interfaces to fully exploit the available bandwidth.
HLS tools can automate this process, but the designer must follow strict coding-style rules. If the bus width is not evenly divisible by the data width (e.g., when using custom-precision data types) or if the arrays are not power-of-two length, the HLS-generated accelerator will likely not fully utilize the available bandwidth, demanding even more manual effort from the designer.
We propose a methodology to automatically find and implement a data layout that, when streamed between memory and an accelerator, uses a higher percentage of the available bandwidth than a naive or HLS-optimized design. 
We borrow concepts from multiprocessor scheduling to achieve such high efficiency.
\end{abstract}


\maketitle


\section{Introduction}

Optimizing data transfers is one of biggest challenges in computing today \cite{Dally2020, Shalf2020}. Many applications, particularly big data and machine learning (ML) algorithms, require huge amounts of data to be transferred and often this is an extreme bottleneck~\cite{9473940}. 
A lot of effort has been put into optimizing the computational aspects of these algorithms, particularly in the development and improvement of high-level synthesis (HLS) tools~\cite{Ferrandi2021}. 
However, the speedup gained on the computation side has not been matched on the data-movement side and thus these applications cannot take full advantage of the optimized accelerators.
In an attempt to solve this problem, High-Bandwidth Memory (HBM) architectures with wide data busses are increasingly common. The Xilinx Alveo u280 has HBM with a maximum bandwidth of 460~GB/s and the Intel Stratix 10 MX has HBM with a maximum bandwidth of 409~GB/s. However, it is very difficult to realistically achieve these high bandwidths. Designers must put in a lot of manual effort to carefully ensure their design utilizes the full bus every single clock cycle~\cite{Choi2020}. In even the simplest designs, this data orchestration can be quite complex and resource intensive. 
In some cases, HLS tools can automatically fill the wide bus by unrolling the arrays, but only if the design meets stringent requirements. For instance, the bus width should be evenly divisible by the data width and the array length should be a power of two. 
Designers can manually make adjustments to meet these requirements or put in more effort to manually pack the bus. Even more effort is needed for a highly custom solution beyond packing equally sized data into evenly divided ``lanes''. 
These highly custom designs are not uncommon, especially with custom-precision data types increasingly used in ML applications \cite{Mahdi2018}. These arbitrarily-sized data prove difficult to fit onto a fixed-width bus.

We propose \textbf{Iris}, an algorithm for automatically finding a \textit{data layout} (i.e., an organization of data in memory and in the bus lanes) such that, when streamed to an accelerator, maximizes the use of the available bandwidth. We borrow concepts from processor scheduling to solve this problem.
Our contributions are: 
\begin{itemize}[leftmargin=*,topsep=3pt]
    \item An algorithm to automatically find an efficient data layout;
    \item A methodology for generating a host-side function that creates a unified array of all the input data in the specified layout; 
    \item A methodology for generating the accelerator-side, HLS-ready modules to convert such data layouts into streams for the kernels. 
\end{itemize}
The automation of this process is useful not only for reducing manual designer effort, but also for rapid design-space exploration while tuning the width of custom-precision data types.

\section{Related Work on HBM Challenges}

With the introduction of HBM architectures, designers should follow recommended guidelines to exploit the increased bandwidth. First, they should use a data width and clock frequency compatible with the architecture. For instance, the HBM in the Xilinx Alveo u280 platform operates at 450~MHz with a data width of 256~bits per channel, so the design should either use this frequency and width or 225~MHz with a 512~bit width. Additionally, transactions should be as large as possible to minimize the overhead per transaction~\cite{Soldavini2022}.

Several works focused on maximizing the use of the channel bandwidth. The work in \cite{Shi2021} analyzes database applications accelerated on FPGAs with HBM. They explicitly craft their designs to ensure that queries return parallelized data to use the full bandwidth. 
The work in \cite{Choi2021} proposes HBM Connect, a customized crossbar for HBM access. They use a virtual HLS FIFO buffer to gather data for read and write operations. On the Alveo u280, they achieve up to 185 GB/s over 16 channels (where the ideal bandwidth would be 230 GB/s). 
The work in \cite{Hogervorst2021} proposes a novel sparse matrix-vector multiplier and an ILU0 preconditioned BiCGStab solver. They explicitly design their computation pipeline to accept a full 512-bit cache line. The main focus is to rearrange the data access (and therefore the compute kernel) to avoid transferring zero data. 
An accelerator for the single-source shortest path problem is proposed in \cite{Chi2022}. Due to the random-access nature of graph problems, they focus on improving throughput and optimizing for HBM.
The work in \cite{Cong2017} proposes an automated methodology for HLS kernels to use more bandwidth than their naive code. Their methodology trades off between using more BRAM or achieving a greater speed.

Since bandwidth is valuable, minimizing the amount of data transferred is important. 
The deep CNN design in \cite{Shah2018} aims at exploiting DRAM bandwidth, reducing the number of data transfers and avoiding stalls. 
The stream analytics engine in \cite{Miao2019} reduces the amount of data to be accessed by putting a smaller, more regularly accessed portion of the data (pointers) into HBM and uses those to reduce DRAM accesses.
The LLVM pass in \cite{Khaldi2016} partitions data between DRAM and HBM. It could be extended to LLVM-based HLS tools. Deciding where to store data can help relieve some bandwidth congestion.
Using a custom-width data format can reduce the total data by reducing the bit-width of each element. 
Custom precision was used in \cite{Ballard2021} to accelerate neural network training. This work targeted only CPU with no memory considerations. 
To the best of our knowledge, no prior work has focused on bandwidth optimization with custom data types.
\begin{table}[tbp]
    \centering
    \caption{Summary of notation used in this work}\label{tab:notationsummary}
    \scalebox{\tabscale}{
    \begin{tabular}{@{}rl@{}}\toprule
        $m$ & There are $m$ processors \\ 
        $j$ & There are $j$ tasks \\
        $\delta_j$ & The maximum number of processors task $j$ can use at once \\ 
        $d_j$ & The due date of task $j$ (time when $j$ would ideally finish) \\
        $r_j$ & The release time of task $j$ (earliest time a task can start)\\ 
        $C_j$ & Completion time of task $j$ \\
        $C_{max}$ & Makespan, maximum completion time of all tasks \\
        $L_j$ & Lateness ($C_j - d_j$) of task $j$ \\ 
        $L_{max}$ & Maximum lateness of all tasks \\
        \bottomrule
    \end{tabular}
    }
\end{table}

\section{Problem Formulation as Scheduling}\label{sec:scheduling}

Scheduling is a popular research problem to decide how jobs are assigned to resources to reduce the overall time to complete all activities while satisfying the constraints~\cite{Sahni76}. 

In our case, given a bus width ($m$) and a set of accelerator arrays, each with bitwidth ($W_j$), depth ($D_j$), and desired due date ($d_j$), we want a memory layout where data are packed most densely and the arrays arrive as close to their due dates as possible when transferred from memory to the accelerator. This can be viewed as a processor scheduling problem as follows: an $m$-bit wide bus is a multiprocessor system made of $m$ identical processors and the data arrays are the tasks, $j$, with due dates $d_j$, and processing times $p_j = W_j\times D_j$. These ``tasks'' are preemptible, i.e., they can be scheduled discontinuously without incurring additional overhead. The ``tasks'' will be scheduled on multiple processors at once. The maximum number of bits an array can use on the bus at a time, or the maximum number of processors a task can use at once, is notated by $\delta_j$.
Arrays may be needed at different times in an accelerator. So each has a due date $d_j$, derived from the dataflow graph and the latencies of the nodes. To ensure the arrays can arrive as shortly after their due date as possible, we use the maximum lateness, $\gamma = L_{max} = \max_j{(L_j)}$, optimality criterion, where $L_j = C_j - d_j$ is the lateness of an array and $C_j$ is the completion time, or last cycle the array is on the bus. All together, our problem is as follows: {\em \uline{in a system with $m$ identical processors, we want to schedule preemptible tasks across several processors (where task $j$ gains linear speedup by being scheduled on up to $\delta_j$ processors at once) such that each task is finished as soon after its due date $d_j$ as possible}}. 
\begin{table}[tbp]
    \centering
    \caption{Summary of additional symbols used in algorithms}
    \label{tab:algsymbsummary}
    \scalebox{\tabscale}{
    \begin{tabular}{@{}rl@{}}\toprule
        $p_j$ & Processing time (time units needed to execute) task $j$\\
        $p_{tot}$ & Total processing time for all tasks (i.e. $C_{max}$ if $m=1$) \\
        $l$ & Number of unique due dates $d_j$ \\
        $d_{max}$ & Maximum (latest) due date of all $d_j$ \\
        $W_j$ & Bitwidth of $j$ \\
        $B_{eff}$ & Bandwidth efficiency \\
        $h(j)$ & Height (minimum possible execution time) of $j$ \\
        $R_k$ & Set of tasks with release time $r_k$ \\
        $\beta_j$ & Processors allocated to $j$ \\
        $t$ & Current timestep being processed \\
        $\tau$ & Length of interval being scheduled \\
        \bottomrule
    \end{tabular}}
\end{table}

\section{Iris: Our Data Layout Algorithm}\label{sec:layoutsolving}

\begin{figure}
    \centering
    \includegraphics[width=0.9\columnwidth]{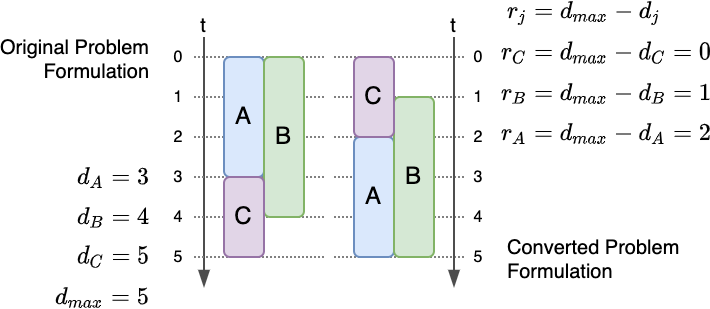}
    \caption{Sample schedule showing conversion between due dates and release times.}
    \label{fig:rj}
\end{figure}

We can find an $O(n^2)$ solution to an isomorphic problem of our formulation 
in \cite{Drozdowski1996}. This isomorphic problem uses release times $r_j$, or the time step when a task $j$ is ready to begin execution, instead of due dates $d_j$, and optimizes the completion time, $C_{max} = \max_j(C_j)$ (also known as schedule length or makespan), instead of the maximum lateness $L_{max}$. This problem is described as follows:
in a system with $m$ identical processors, we want to schedule preemptible tasks with release time $r_j$ across several processors to minimize the total schedule length ($C_{max}$).
To convert between the two problems, each due date $d_j$ is converted to a release time $r_j$ by subtracting it from the maximum (latest) due date of all tasks such that $r_j = d_{max} - d_j$. Also, the solution schedule to the isomorphic problem should be read backward to find the solution to the original problem. In this way, tasks that originally have the latest due dates will have the earliest release times in the ``backward'' schedule. \autoref{fig:rj} shows how converting due dates $d_j$ into release times $r_j$ can yield the same schedule but reversed in time.
\algrenewcommand\algorithmicindent{0.6em}
\renewcommand{\thealgorithm}{1.\arabic{algorithm}}
\setcounter{algorithm}{0}
\begin{algorithm}[bp]
\small
\caption{Layout Algorithm}\label{alg:top}
\begin{algorithmic}[1]
\State $t := 0$
\State Group tasks with release time $r_k$ in set $R_k$, $k=1,\dotsc,l$ \label{alg:top:Rk}
\For{$k := 1$ to $l$}
    \State Order $R_k$ by nonincreasing values of $h(j)$ \label{alg:top:Rksort}
    \While {$(r_{k+1}<t)$ \textbf{and} $(\exists_{j\in R_k} h(j) > 0)$}\label{alg:top:while}
        \State \Call {Find\_Capabilities}{$R_k$, $\overline{\beta}$}
        \If{$\exists_{j,j+1\in R_k}h(j) > h(j+1)$}
            \Statex \Comment{Shortest time before $h(j), h(j+1)$ are equal}
            \State $\tau' := \min{\left\{%
            \vcenter{\hbox{$\displaystyle
            \frac{h(j) - h(j+1)}{\frac{\beta_j}{\delta_j}-\frac{\beta_{j+1}}{\delta_{j+1}}}%
            $}}
            :%
            \frac{\beta_j}{\delta_j}\neq\frac{\beta_{j+1}}{\delta_{j+1}},h(j) > h(j+1)%
            \right\}
            }$ \label{alg:top:taup}
        \Else 
            \State $\tau' := \infty$
        \EndIf
        \State $\tau'' := h(|R_k|)$ \Comment{Time to earliest completion of any task}
        \State $\tau := \min{\{\tau', \tau'', r_{k+1}-t\}}$ \label{alg:top:taurk}\Comment{Interval is until next change}
        \State For all $j\in R_k$, schedule $j$ on $\beta_j$ processors for the interval $[t,t+\tau]$ \label{alg:top:sched}
        \State $h(j) := h(j) - \frac{\tau\beta_j}{\delta_j}$ for $j\in R_k$ \Comment{Subtract this proc. time from $h$}
        \State $t=t+\tau$ \Comment{Update the timestep}\label{alg:top:tupdate}
    \EndWhile
    \Statex \Comment{Add any unfinished tasks to the next batch, $R_{k+1}$}
    \If {$\exists_{j\in R_k}h(j)>0$} $R_{k+1} := R_{k+1}\cup \{j:j\in R_k, h(j) > 0\}$ 
    \EndIf
\EndFor
\algstore{beforeFC}
\end{algorithmic}
\end{algorithm}
\begin{algorithm}[tbp]
\caption{Find\_Capabilites Procedure}\label{alg:fc}
\small
\begin{algorithmic}[1]
\algrestore{beforeFC}
\Procedure{Find\_Capabilities}{$X$, $\overline{\beta}$} \Comment{$X$ is a set of tasks}
    \State $\overline{\beta} := \overline{0}$ \Comment{$\overline{\beta}$ is a vector of \# processors allocated to each task $j$}
    \State $avail := m$ \Comment{$avail$ is the number of free processors}
    \While{$avail > 0$ \textbf{and} $|X| > 0$} 
        \State $T := $ set of the highest tasks in $X$ with $h(j) > 0$ \label{alg:fc:T}
        \If{$\sum_{j\in T}\delta_j > avail$}
            \State $\beta_j := \Call{LRM\_Allocation}{T}$ \label{alg:fc:ham}; $avail := 0$
        \Else \Comment{Tasks in $T$ can use at most $avail$ processors}
            \State $\beta_j := \delta_j$ for $j\in T$;
            $avail := avail-\sum_{j\in T}\delta_j$ \label{alg:fc:elsebody}
        \EndIf
        \State $X := X - T$ \label{alg:fc:XminT} \Comment{Remove scheduled tasks from the working set, $X$}
    \EndWhile
\EndProcedure
\algstore{beforeLRM}
\end{algorithmic}
\end{algorithm}

\begin{algorithm}[tbp]
\caption{LRM\_Allocation Procedure}\label{alg:lrm}
\small
\begin{algorithmic}[1]
\algrestore{beforeLRM}
\Procedure{LRM\_Allocation}{$T$}
    \State $quota := \left(\sum_{j\in T}\delta_j\right) / avail$ \Comment{Hare quota of processors}
    \For{$j\in T$}
        \State $v_j := \frac{\delta_j}{quota}$ \Comment{$v_j$ is the processors requested per $quota$}
        \State $\beta_j := \left\lfloor{{v_j}/{\delta_j}}\right\rfloor$ \label{alg:lrm:floor} \Comment{Assign $\beta_j$ the largest multiple of $\delta_j$ below $v_j$}
        \State $rem_j := v\mod\ \delta_j$ \Comment{Keep track of the remainder}
        \State $avail := avail - \beta_j$
    \EndFor
    \State Sort $T$ by decreasing $rem_j$
    \For{$j\in T$}
        \Statex \Comment{If $j$ fits in the remaining space, schedule it}
        \If{$avail > W_j$}
            $\beta_j := \beta_j + 1$; $avail := avail - 1$ 
        \EndIf
        \If{$avail = 0$}
            \textbf{return} \Comment{When there is no more space, done}
        \EndIf
    \EndFor
\EndProcedure
\end{algorithmic}
\end{algorithm}



Algorithm~\ref{alg:top} shows our proposed layout finding algorithm, called \textit{Iris}. In Greek mythology, Iris is the messenger of the gods. Additional symbols used in the algorithms are summarized in \autoref{tab:algsymbsummary}.
The algorithm in \cite{Drozdowski1996} is designed to optimize the completion time, $C_{max}$, given a set of release times, $r_j$, when tasks are available to be processed. The algorithm can be converted to optimizing for the minimum lateness, $L_{max}$, by changing a set of $l$ due dates, ${d_1 \leq d_2 \leq \dotso \leq d_l}$, to release times as follows: $r_j = d_{max} - d_j$. 
After scheduling using these new release times, the schedule can be read backward to optimize for $L_{max}$ for input arrays using the due dates. 

To adapt this algorithm for the \textbf{bus layout problem}, we modified it as follows. 
Instead of using a simple ratio, Iris uses the largest-remainder method (also known as the Hamilton Method) of apportionment to allocate processors to tasks \cite{Kohler2012}. 
This method ensures tasks are assigned whole numbers of ``processors'' (i.e., bus bit lanes). Also, 
regular multiprocessor tasks can be split arbitrarily, but
array elements are indivisible. For instance, an array with 17-bit elements can use 17, 34, or 51 bits of a 64-bit bus (i.e., transferring one or more elements), but not 20 bits (i.e., transferring parts of the elements).
To schedule indivisible elements, we modified the largest-remainder method to only allocate in multiples of the bitwidth (Line~\ref{alg:lrm:floor}). The remainders can then be greater than one, but always less than the bitwidth of the element. Without this modification, 
the overhead for organizing the data with logical-shift and bitwise operations would be prohibitive.

\begin{table}[tbp]
    \vspace{4pt}\centering
    \caption{Example set of inputs}\label{tab:ex}%
    \scalebox{\tabscale}{
    \begin{tabular}{@{}ccccc@{}} \toprule
        Array & Width ($W$) & Depth ($D$) & Due Date ($d$) & Processing Time ($p=W\times D$) \\
        \midrule
        $A$ & 2 & 5 & 2 & 10 \\
        $B$ & 3 & 5 & 6 & 15 \\
        $C$ & 4 & 3 & 3 & 12 \\
        $D$ & 5 & 4 & 6 & 20 \\
        $E$ & 6 & 2 & 3 & 12 \\\bottomrule
    \end{tabular}
    }
\end{table}

A small example is presented here. \autoref{tab:ex} lists the characteristics of arrays.
The total processing time, $p_{tot}$ (total number of bits in all of the arrays) is 69. 
Ideally, $C_{max} \times m$ is as close to $p_{tot}$ as possible to ensure there is the smallest amount of wasted bandwidth. Therefore, we compute \textbf{bandwidth efficiency} as:
\begin{equation}
B_{eff}=\dfrac{p_{tot}}{C_{max}\times m}
\end{equation}
So, the ideal case is a value of 1 (or 100\%), which means that the accelerator is fully utilizing the bandwidth. 

\begin{figure*}
\begin{minipage}[t]{0.3\textwidth}
    \centering
    \includegraphics[scale=\schedscale]{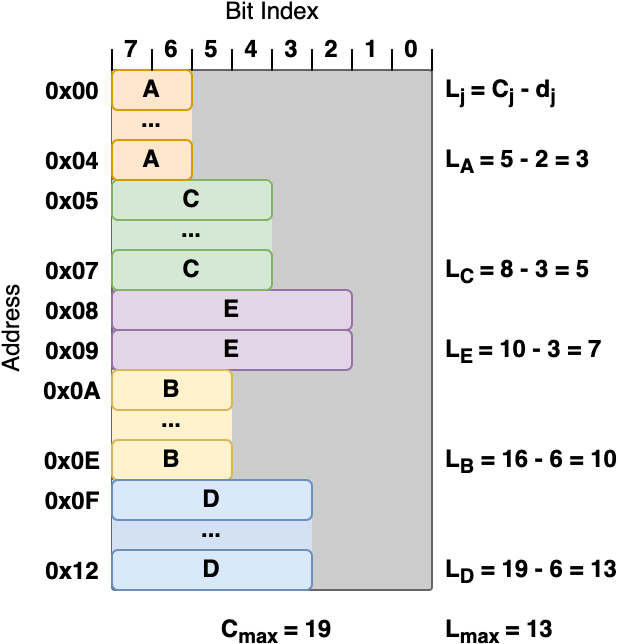}
    \caption{A naive layout for the example arrays.}
    \label{fig:naive}
\end{minipage}\hfill
\begin{minipage}[t]{0.3\textwidth}
    \centering
    \includegraphics[scale=\schedscale]{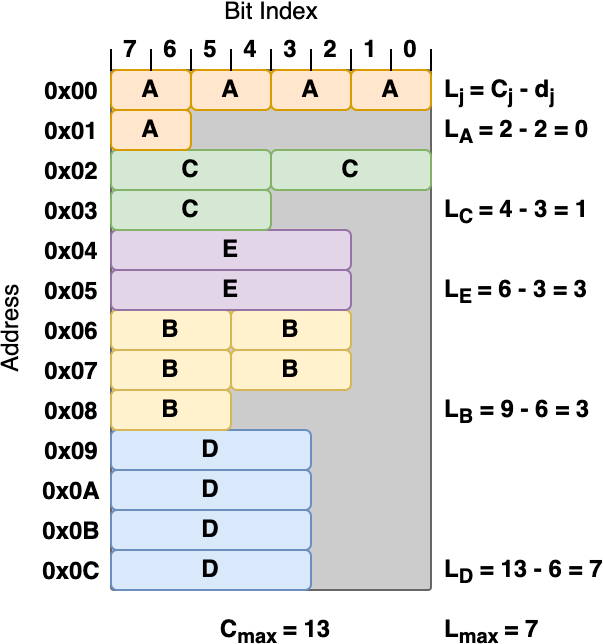}
    \caption{A homogeneously packed naive layout for the example arrays.}
    \label{fig:naivepack}
\end{minipage}\hfill
\begin{minipage}[t]{0.3\textwidth}
    \centering
    \includegraphics[scale=\schedscale]{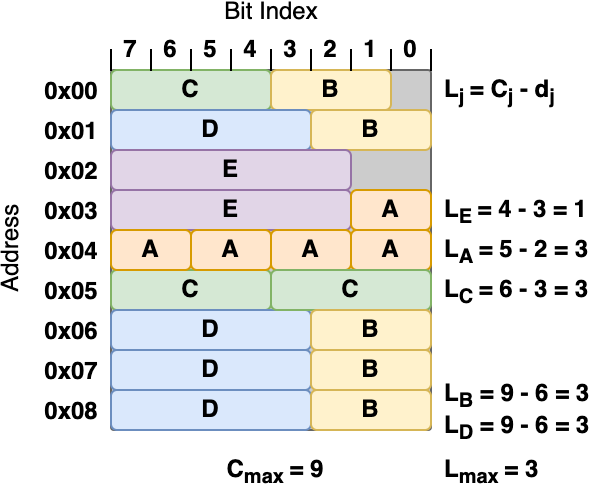}
    \caption{The layout for the example arrays generated by our method.}
    \label{fig:ours}
\end{minipage}
\vspace{-10pt}
\end{figure*}

A completely naive method would be to sort the arrays by increasing due date and place one element of each array into each 8-bit slot of memory. The resulting diagram is shown in \autoref{fig:naive}. 
Array $D$ would arrive 13 cycles after its due date of $d_D = 6$ ($L_{max}=13$). The efficiency of this layout is $\frac{69}{19 \times 8}=45.4\%$. 
An improvement would be to pack as many elements of an array as possible onto the bus at once. This homogeneous packing is more dense but still fairly naive. This layout is shown in \autoref{fig:naivepack}.
In this layout, $L_{max} = L_D = 7$ and the efficiency is $\frac{69}{13 \times 8} = 66.3\%$. 

\begin{table}[tbp]
    \vspace{4pt}\centering
    \caption{$r_j$, $\delta_j$, and $h(j)$ for each array. Arrays sorted by nondecreasing $d_j$. $d_{max} = \max_j(d_j)$}\label{tab:rj}%
\scalebox{\tabscale}{
\begin{tabular}{crrrrr} \toprule
        Array & $A$ & $C$ & $E$ & $B$ & $D$ \\ 
        $j$   & 1   &  2  & 3   & 4   & 5 \\\midrule
        $d_j$ & 2   & 3   & 3   & 6   & 6   \\
        $r_j$ & 4 & 3 & 3 & 0 & 0 \\
        $\delta_j$ & 8 & 8 & 6 & 6 & 5 \\\midrule
        $h(j)$ & 2 & 2 & 2 & 3 & 4 \\\bottomrule
    \end{tabular}}\vspace{4pt}
\end{table}

\begin{figure}
    \centering
    \includegraphics[scale=\schedscale]{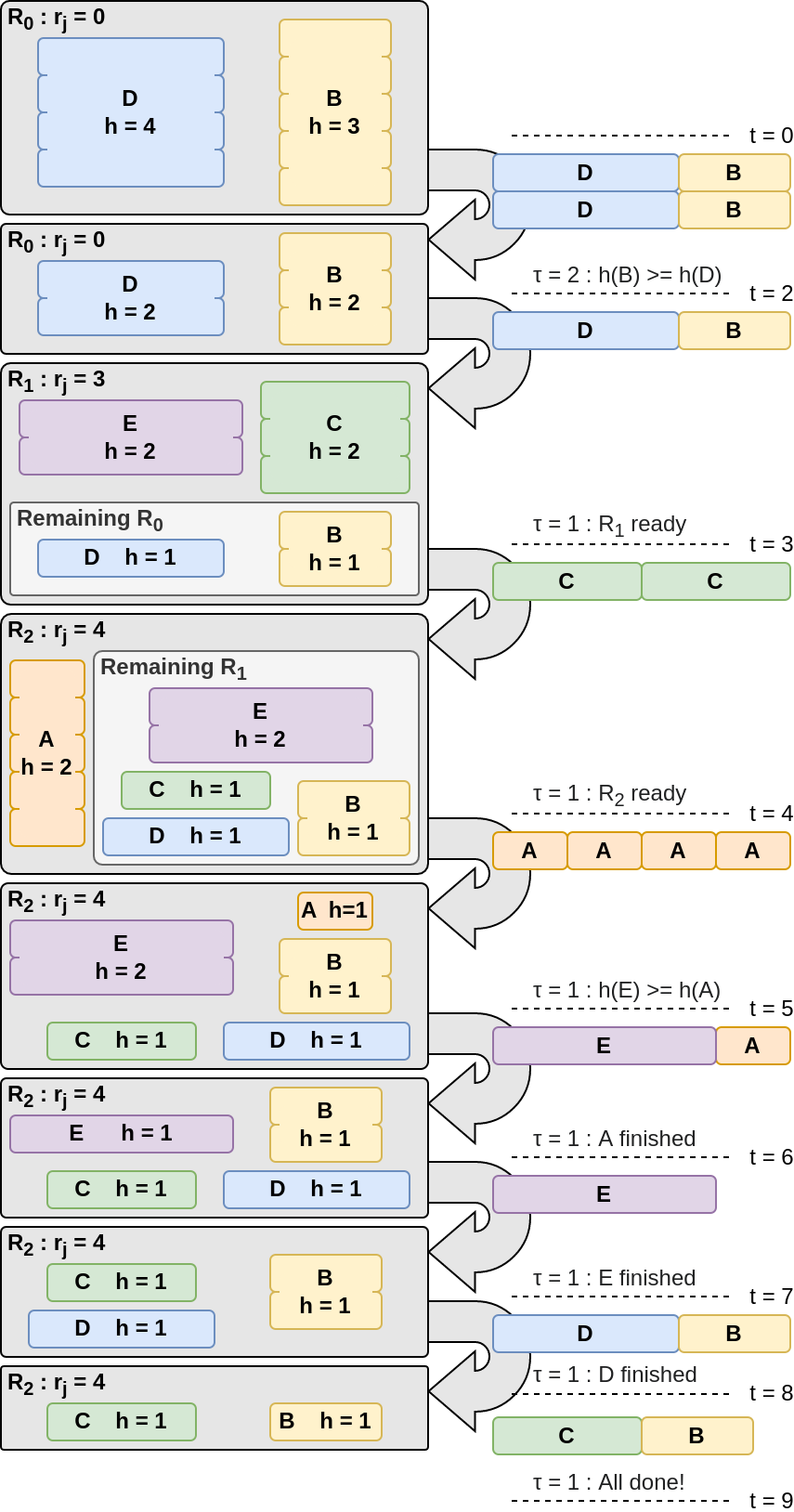}
    \caption{Our process of ``scheduling'' arrays into the layout.}
    \label{fig:alg}\vspace{2pt}
\end{figure}

To convert this problem into one the algorithm can solve, the release times $r$ should be computed from the set of due dates $d$ as $r_j=d_{max}-d_j$. We show this and the computations for the maximum bits per cycle for an array, $\delta_j=\left\lfloor m /W_j\right\rfloor\times W_j$, and for the heights, $h(j) = {p_j}/{\delta_j}$ in \autoref{tab:rj}.
The set of unique release times is $r = \{0, 3, 4\}$. Using this set, the arrays must be sorted into groups, $R_k$, based on $r_k$ (Line~\ref{alg:top:Rk}). Within each $R_k$, the arrays are ordered by nonincreasing values of $h(j)$: $R_0 = \{D, B\}$ ($r_j=0$), $R_1 = \{C, E\}$ ($r_j=3$), and $R_2 = \{A\}$ ($r_j = 4$). Then, each $R_k$ is processed in order. The algorithm executed on the example arrays is shown in \autoref{fig:alg}.

Each large box on the left side of \autoref{fig:alg} shows the current working group, $R_k$, and the current ready time, $r_j$, in the top left corner. Inside this box is each array currently ready to be processed. The curved arrows indicate the end of an iteration of the while loop (Line~\ref{alg:top:while}) of Algorithm~\ref{alg:top}. 
The reason for the value of $\tau$ in that iteration is listed on the arrow on the right side along with which array elements get placed into the layout at that interval. When the next $R_k$ group is ready, the arrays which are not yet fully processed are added to the new group. At the end of the procedure, the final layout must be reversed to target $L_{max}$, as shown in \autoref{fig:ours}. 
The latest arrays arrive only 3 cycles after their due dates ($L_{max} = 3$). The efficiency is now $\frac{69}{9\times 8} = 95.8\%$, wasting only 3 bandwidth bits. 

\section{Code Generation}\label{sec:codegen}

Because all array details are statically known, we execute Iris during the compilation part to determine the data layouts and generate the necessary functions for decoding them into the accelerator. 

\vspace{4pt}
\noindent\textbf{Host-Side Organization. }
To transfer data from the CPU using the proposed layout, the host must aggregate the arrays into the layout efficiently. 
The procedure for organizing the data, given pointers to all of the input arrays and a pointer to allocated memory the size of the layout ($m \times C_{max}$), is as follows. 
We create each layout cycle using the machine-word-size of the host. For example, if the layout is for a 256-bit bus and the host uses a 64-bit word size, we organize the memory line in four adjacent \texttt{uint64} elements. 
The generator iterates over each array assigned to each cycle and logical-shift-left the next element of that array into the current word. When this word is full, it places it in its appropriate memory location and starts the next one. After placing each array element, it increments its array pointer such that the next element will be inserted in the next steps. When an element spans across words, it shifts in the remaining bits to the top of the next word.

\lstset{basicstyle=\footnotesize\ttfamily,breaklines=true}
\begin{lstfloat}
    \centering
\begin{lstlisting}[language=C]
void pack(int* A, int* B, int* C, int* D, int* E, 
          unsigned char* out) {
    unsigned char curr;
    // 0 : C, B
    curr  = ((*C++) & C_MASK) << (B_WIDTH + 1);
    curr |= ((*B++) & B_MASK) << (1); 
    *out++ = curr;
    // 1 : D, B
    curr  = ((*D++) & D_MASK) << (B_WIDTH);
    curr |= ((*B++) & B_MASK);
    *out++ = curr;
	// ...
    for (unsigned int t = 0; t < 2; t++) {
	// 7-8 : D, B
	curr  = ((*D++) & D_MASK) << (B_WIDTH);
	curr |= ((*B++) & B_MASK);
	*out++ = curr;
    }
}
\end{lstlisting}
    \caption{Sample host function for organizing arrays into the layout}
    \label{lst:host}
\end{lstfloat}

The \texttt{C} function for the data organization of the example in \autoref{sec:layoutsolving} is shown in Listing~\ref{lst:host}.
\texttt{X\_WIDTH} and \texttt{X\_MASK} constants represent the width of the array and a bitmask of the appropriate width, respectively. 
\texttt{(*X++)} will get the value at pointer \texttt{X}, and then post-increment it. In the case of $\tau>1$ (e.g., in cycles 7-8), we use a \texttt{for} loop to create the same layout over several cycles. This simple function can be automatically generated from the given layout. 

\begin{lstfloat}[t]
    \centering
\begin{lstlisting}[language=C]
#define BUSWIDTH 8
#define A_FIFO_DEPTH 3
#define C_FIFO_DEPTH 1
void read_data(ap_uint<BUSWIDTH>* in_buf, 
               hls::stream<ap_uint<A_WIDTH>>& dataA, 
	           // ...
               hls::stream<ap_uint<E_WIDTH>>& dataE) {
    ap_uint<BUSWIDTH> elem;
    dataA_t tmpA[A_FIFO_DEPTH];
    dataC_t tmpC[C_FIFO_DEPTH];
    for(unsigned int t = 0; t < 9; t++){
#pragma HLS pipeline II=1
        elem = in_buf[t];
        if (t == 0) {
            dataC << elem.range(7, 4);
            dataB << elem.range(3, 1);
        } else if (t == 1) {
            dataD << elem.range(7, 3);
            dataB << elem.range(2, 0);
                // ...
        } else if (t >= 7 && t <= 8) {
            dataA << tmpA[0];
            dataD << elem.range(7, 3);
            dataB << elem.range(2, 0);
        }
    }
}
\end{lstlisting}
    \caption{Sample HLS module for a data read module to decode the layout for an accelerator (Trimmed for brevity)}
    \label{lst:hls}
\end{lstfloat}

\vspace{4pt}
\noindent\textbf{Accelerator-Side Decoding. }
Once the data are transferred into memory accessible by the accelerator (e.g., Xilinx HBM), the accelerator must interpret the data. We implement specialized modules to exchange data between memory and the appropriate streams. 

The data-read module must have an initiation interval of~1 to maintain maximum bandwidth utilization. To achieve this, enough local memory ports must be available to store all data elements on the bus at once. For data elements that only appear once in any cycle of the layout, the stream interface or a private local memory (PLM) is sufficient. However, if two or more elements from the same array are present in a single cycle, we need extra memories to temporarily store these elements to free up the bus quickly, rather than waiting for several cycles to read each element off the bus. For instance, if at most four elements from array \texttt{A} are on the bus in a single cycle, we need four write ports~\cite{Pilato2017}. This can be implemented as a three-element shift-register where \texttt{A[i]} is written straight to the destination, and \texttt{A[i+1]}, \texttt{A[i+2]}, and \texttt{A[i+3]} are parallel-loaded into the shift-register, and each successive cycle has the next element written to the destination. However, if more elements of \texttt{A} are on the bus in these three cycles, additional depth might be needed in the shift-register. 
The maximum depth of the shift-register for an array is determined during layout creation by a running sum over each schedule interval.  

Listing~\ref{lst:hls} shows a sample read module for the example layout written in Xilinx-style HLS code, using the HLS library for \textit{arbitrary Precision Types} (\texttt{ap\_uint}).
This module sends each element to a stream for the appropriate array. Downstream dataflow modules can begin execution as soon as the first elements are sent. The constant \texttt{X\_WIDTH} values are also the bitwidths of their respective arrays. The HLS tool estimates a latency of 11 clock cycles with only 29 flip-flops and 194 LUTs. For the naive read module (\autoref{fig:naive}) HLS estimates a latency of 43 cycles and uses 54 flip-flops and 452 LUTs. Thus, we improve both latency and resource requirements.

\section{Evaluation}
We implemented a prototype of Iris in Python which receives the input (e.g., bus bitwidth and array details) as a JSON file. This file can be automatically generated by reading array details from the kernel during HLS.
To evaluate our method, we analyze several layouts generated for two real accelerators. In all cases, we use $m=256$ to target the real bus width of the HBM on the Alveo u280.

\vspace{4pt}\noindent\textbf{Inverse Helmholtz. } 
The work in \cite{Soldavini2022} aims at deploying the Inverse Helmholtz operator on the Alveo u280. This operator is the building block of a computational fluid dynamics application. Due to the physical nature of the values, each array element uses 64 bits (\texttt{double}). In~\cite{Soldavini2022}, the authors examine different strategies for optimizing the data transfers but using the \textit{packed naive} approach for the HBM. \autoref{tab:inputs} shows the depths and due dates of each array.
$d_S$ and $d_u$ are simply the earliest time by which these arrays can feasibly be finished. $D$ is needed later than $u$ and $S$, so $d_D$ is the earliest time by which $u$ and $S$ could both be feasibly finished by.

\begin{table}[tbp]
    \centering
    \caption{Set of inputs for our test accelerators}\label{tab:inputs}%
\scalebox{\tabscale}{
    \begin{tabular}{cccccc} \toprule
        Accelerator & Array & Width & Depth & Due Date ($d$) \\ \midrule
       \multirow{3}{*}{Inv. Helmholtz} & $u$ & 64 & 1331 & 333 \\
       &  $S$ & 64 & 121  & 31 \\
       &  $D$ & 64 & 1331 & 363 \\\midrule
       \multirow{2}{*}{Matrix Multiplication} & $A$ & 64 & 625 & 157 \\
       & $B$ & 64 & 625 & 157    \\\bottomrule
    \end{tabular}
    }
\end{table}

A naive layout following the pattern in \autoref{fig:naivepack} has an efficiency of 99.8\% and $L_{max} = 364$. Our layout is 99.9\% efficient, using one less cycle, and $L_{max} = 333$. Because these data widths are all evenly divisible into the bus-width, the metrics for our layout are nearly the same. However, we reduce the FIFO depth from 998 for $u$ and $D$ and 90 for $S$ to 666 for $u$ (-33\%), 636 for $D$ (-36\%), and 30 for $S$ (-67\%).
In the naive layout, four elements of each array are nearly always sent on the bus at a time. In our layout, instead, the three arrays are often interleaved together in the same cycle, relieving the contention pressure on the FIFOs. This improvement is important since BRAMs are usually a limiting factor for data-intensive applications. 

\begin{table}[tbp]
    \centering
    \caption{Layout metrics with varied $\delta/W$ (Inv. Helmholtz)}\label{tab:helmres}%
\scalebox{\tabscale}{
    \begin{tabular}{ccccccc} \toprule
         & & & \multicolumn{4}{c}{$\delta/W$} \\
         & & Naive & 4 & 3 & 2 & 1 \\ \cmidrule(rr){1-2}\cmidrule(lr){3-3}\cmidrule(ll){4-7}
        \multicolumn{2}{c}{Efficiency} & 99.8\% & 99.9\% & 98.8\% & 97.9\% & 51.1\% \\
        \multicolumn{2}{c}{$C_{max}$} & 697 & 696 & 704 & 711 & 1361 \\
        \multicolumn{2}{c}{$L_{max}$}  & 364    & 333    & 341    & 348    & 998\\
        \multirow{3}{*}{{\makecell{FIFO\\Depth}}} 
        & $u$ & 998 & 666 & 667 & 665 & 0 \\
        & $S$ & 90  & 30  & 30  & 15  & 0 \\
        & $D$ & 998 & 636 & 631 & 620 & 0 \\
        \bottomrule
    \end{tabular}
    }
\end{table}

We can even vary the maximum number of times an array can have elements on the bus in one cycle by reducing $\delta$ to a lower multiple of the bit-width. \autoref{tab:helmres} summarizes the results when constraining the arrays as such.
For $\delta/W > 1$, we slightly improve FIFO depth as $\delta/W$ decreases, along with slight efficiency and $L_{max}$ degradation. When $\delta/W=1$, the efficiency drops to 51.1\% because there are only 3 arrays, so it is impossible to fill the entire bandwidth if they are all only allowed to have one element on the bus at a time. However, we eliminate the need for extra write-port FIFOs since only one element must be written to any array at a time. If a design is having difficulty due to area constraints, and not having data-transfer bottleneck issues, this layout may be useful. 

\vspace{4pt}\noindent\textbf{Matrix Multiplication. }
We also test layouts for Matrix Multiplication, which is popular in many tensor-based applications \cite{Haidar2018, BenNun2019}. Its inputs are summarized in \autoref{tab:inputs}.
The due dates for this application are both as soon as possible, as both inputs are needed at the same time. With $W=64$ again, the naive layout and our layout perform nearly identically, with our layout only having slightly better $L_{max}$ and FIFO depth. However, when we vary the bitwidths with custom precision data types, Iris achieves better results. Results for the matrix multiply layout are summarized in \autoref{tab:mmres}.

With custom precision data types, it is difficult to fit neatly into the bus width. Our layout algorithm determines how to better utilize the bandwidth without sacrificing performance. %
In the case of 64-bit data, the schedule length is reduced by one cycle, but the memory resources are reduced by 33\%. 
For 33- and 31-bit widths, the schedule length, which directly correlates to data transfer time, is reduced by 5\% and the overall FIFO memory resources are reduced by 13\%.
Finally, for the 30- and 19-bit width, the schedule length is reduced by 2\% and the memory resources are reduced by 8\%.

\begin{table}[tbp]
    \centering
    \caption{Layout metrics with varied $W$ (Matrix Multiply)}\label{tab:mmres}%
\scalebox{\tabscale}{
    \begin{tabular}{cccccccc} \toprule
        \multicolumn{2}{c}{$(W_A, W_B)$} & \multicolumn{2}{c}{(64, 64)}  & \multicolumn{2}{c}{(33, 31)} & \multicolumn{2}{c}{(30, 19)} \\
         & & Naive & Iris & Naive & Iris & Naive & Iris  \\ \cmidrule(rr){1-2}\cmidrule(lr){3-4}\cmidrule(lr){5-6}\cmidrule(lr){7-8}
        \multicolumn{2}{c}{Efficiency} & 99.5\% & 99.8\% & 92.5\% & 98.9\% & 93.5\% & 97.3\%  \\
        \multicolumn{2}{c}{$C_{max}$} & 314 & 313 & 236 & 225 & 206 & 201 \\
        \multicolumn{2}{c}{$L_{max}$}  & 157    & 156 &  79 & 68  & 49 & 44  \\
        \multirow{2}{*}{{\makecell{FIFO\\Depth}}} 
        & $A$ & 468  & 312 & 535 & 467 &    546 & 502  \\
        & $B$ & 468  & 312 & 546 & 478 &    576 & 532  \\
        \bottomrule
    \end{tabular}
    }
\end{table}

\section{Conclusion}
This work presented \textit{Iris}, an algorithm designed to automatically create an efficient data layout that maximizes the use of the available bandwidth. Iris was able to achieve higher bandwidth efficiency and lower lateness $L_{max}$ for various accelerators. Also, the solutions created by Iris use less FPGA resources for the data read module, particularly in the case of the data FIFOs necessary to read from the bus every cycle. Also, as Iris is an automatic process, this relieves the designer of a huge manual effort and can even support rapid design space exploration when using custom data types.

\begin{acks}
This work was partially funded by the EU Horizon 2020 Programme under grant agreement No 957269 (EVEREST).
\end{acks}


\renewcommand{\bibfont}{\footnotesize}
\printbibliography

@article{Shi2021,
author = {Shi, Runbin and Kara, Kaan and Hagleitner, Christoph and Diamantopoulos, Dionysios and Syrivelis, Dimitris and Alonso, Gustavo},
title = {Exploiting {HBM} on {FPGAs} for Data Processing},
year = {2021},
publisher = {Association for Computing Machinery},
journal = {ACM TRETS},
month = {10}
}

@article{Sahni76,
author = {Sahni, Sartaj K.},
title = {Algorithms for Scheduling Independent Tasks},
year = {1976},
issue_date = {Jan. 1976},
publisher = {Association for Computing Machinery},
volume = {23},
number = {1},
journal = {J. ACM},
month = {1},
pages = {116–127},
}

@misc{Choi2020,
  author = {Choi, Young-kyu and Chi, Yuze and Wang, Jie and Guo, Licheng and Cong, Jason},
  title = {When {HLS} Meets {FPGA} {HBM}: Benchmarking and Bandwidth Optimization},
  publisher = {arXiv},
  year = {2020},
  copyright = {arXiv.org perpetual, non-exclusive license}
}

@inproceedings{Choi2021,
author = {Choi, Young-kyu and Chi, Yuze and Qiao, Weikang and Samardzic, Nikola and Cong, Jason},
title = {{HBM} Connect: High-Performance {HLS} Interconnect for {FPGA} {HBM}},
year = {2021},
booktitle = {FPGA},
pages = {116–126}
}

@INPROCEEDINGS{Ferrandi2021,
  author={Ferrandi, Fabrizio and Castellana, Vito Giovanni and Curzel, Serena and Fezzardi, Pietro and Fiorito, Michele and Lattuada, Marco and Minutoli, Marco and Pilato, Christian and Tumeo, Antonino},
  booktitle={DAC}, 
  title={Invited: Bambu: an Open-Source Research Framework for the High-Level Synthesis of Complex Applications}, 
  year={2021},
  volume={},
  number={}
  }

@article{Hogervorst2021,
author = {Hogervorst, Tom and Nane, R\u{a}zvan and Marchiori, Giacomo and Qiu, Tong Dong and Blatt, Markus and Rustad, Alf Birger},
title = {Hardware Acceleration of High-Performance Computational Flow Dynamics Using High-Bandwidth Memory-Enabled Field-Programmable Gate Arrays},
year = {2021},
issue_date = {June 2022},
publisher = {Association for Computing Machinery},
volume = {15},
number = {2},
journal = {ACM TRETS},
month = {12},
articleno = {20}
}

@inproceedings{Miao2019,
author = {Miao, Hongyu and Jeon, Myeongjae and Pekhimenko, Gennady and McKinley, Kathryn S. and Lin, Felix Xiaozhu},
title = {{StreamBox-HBM}: Stream Analytics on High Bandwidth Hybrid Memory},
year = {2019},
booktitle = {ASPLOS},
pages = {167–181}
}

@inproceedings{Chi2022,
author = {Chi, Yuze and Guo, Licheng and Cong, Jason},
title = {Accelerating {SSSP} for Power-Law Graphs},
year = {2022},
booktitle = {FPGA}
}

@ARTICLE{Pilato2017,
    author={Christian {Pilato} and Paolo {Mantovani} and Giuseppe {Di Guglielmo} and Luca P. {Carloni}},
    journal={IEEE TCAD},
    title={System-Level Optimization of Accelerator Local Memory for Heterogeneous Systems-on-Chip},
    year={2017},
    volume={36},
    number={3},
    pages={435-448},
}

@inproceedings{Khaldi2016,
author = {Khaldi, Dounia and Chapman, Barbara},
title = {Towards Automatic HBM Allocation Using LLVM: A Case Study with Knights Landing},
year = {2016},
booktitle = {LLVM-HPC},
pages = {12–20}
}

@inproceedings{Ballard2021,
author = {Ballard, Grey and Weissenberger, Jack and Zhang, Luoping},
title = {Accelerating Neural Network Training Using Arbitrary Precision Approximating Matrix Multiplication Algorithms},
year = {2021},
booktitle = {ICPP Workshops},
articleno = {16},
pages={1--8},
}

@ARTICLE{Shah2018,  author={Shah, Nimish and Chaudhari, Paragkumar and Varghese, Kuruvilla},  journal={IEEE TNNLS},   title={Runtime Programmable and Memory Bandwidth Optimized FPGA-Based Coprocessor for Deep Convolutional Neural Network},   year={2018},  volume={29},  number={12},  
}

@inproceedings{Cong2017,
author = {Cong, Jason and Wei, Peng and Yu, Cody Hao and Zhou, Peipei},
title = {Bandwidth Optimization Through On-Chip Memory Restructuring for {HLS}},
year = {2017},
booktitle = {DAC}
}

@article{Drozdowski1996,
title = {Real-time scheduling of linear speedup parallel tasks},
journal = {Information Processing Letters},
volume = {57},
number = {1},
pages = {35-40},
year = {1996},
author = {Maciej Drozdowski}
}

@INPROCEEDINGS{9473940,
  author={Pilato, Christian and Bohm, Stanislav and Brocheton, Fabien and Castrillon, Jeronimo and Cevasco, Riccardo and Cima, Vojtech and Cmar, Radim and Diamantopoulos, Dionysios and Ferrandi, Fabrizio and Martinovic, Jan and Palermo, Gianluca and Paolino, Michele and Parodi, Antonio and Pittaluga, Lorenzo and Raho, Daniel and Regazzoni, Francesco and Slaninova, Katerina and Hagleitner, Christoph},
  booktitle={DATE}, 
  title={{EVEREST}: A design environment for extreme-scale big data analytics on heterogeneous platforms}, 
  year={2021},
  volume={},
  number={},
  pages={1-6}
  }

@article{Soldavini2022,
author = {Soldavini, Stephanie and Friebel, Karl F. A. and Tibaldi, Mattia and Hempel, Gerald and Castrillon, Jeronimo and Pilato, Christian},
title = {Automatic Creation of High-Bandwidth Memory Architectures from Domain-Specific Languages: The Case of Computational Fluid Dynamics},
year = {2022},
publisher = {Association for Computing Machinery},
address = {New York, NY, USA},
journal = {ACM Trans. Reconfigurable Technol. Syst.},
month = {09}
}

@article{Kohler2012,
author = {Ulrich Kohler and Janina Zeh},
title ={Apportionment Methods},
journal = {The Stata Journal},
volume = {12},
number = {3},
pages = {375-392},
year = {2012}
}

@inproceedings{Mahdi2018,
author = {Nazemi, Mahdi and Pedram, Massoud},
title = {Deploying Customized Data Representation and Approximate Computing in Machine Learning Applications},
year = {2018},
booktitle = {ISLPED}
}

@INPROCEEDINGS{Haidar2018,
author={Haidar, Azzam and Tomov, Stanimire and Dongarra, Jack and Higham, Nicholas J.},
booktitle={SC},   
title={Harnessing {GPU} Tensor Cores for Fast {FP16} Arithmetic to Speed up Mixed-Precision Iterative Refinement Solvers},   
year={2018},  
volume={},  
number={}  
}

@article{BenNun2019,
author = {Ben-Nun, Tal and Hoefler, Torsten},
title = {Demystifying Parallel and Distributed Deep Learning: An In-Depth Concurrency Analysis},
year = {2019},
issue_date = {July 2020},
publisher = {Association for Computing Machinery},
volume = {52},
number = {4},
journal = {ACM Comput. Surv.},
month = {8},
articleno = {65}
}

@article{Dally2020,
author = {Dally, William J. and Turakhia, Yatish and Han, Song},
title = {Domain-Specific Hardware Accelerators},
year = {2020},
issue_date = {July 2020},
publisher = {Association for Computing Machinery},
volume = {63},
number = {7},
journal = {Commun. ACM},
month = {6},
pages = {48–57}
}

@article{Shalf2020,
  year = {2020},
  month = jan,
  publisher = {The Royal Society},
  volume = {378},
  number = {2166},
  author = {John Shalf},
  title = {The future of computing beyond {M}oore's {L}aw},
  journal = {Philosophical Transactions of the Royal Society A: Mathematical,  Physical and Engineering Sciences}
}

\end{document}